\documentclass[slac_one]{revtex4}
\usepackage{graphicx}
\usepackage{fancyhdr}
\newcommand{\Eslash}{\mbox{$\rm E \kern-0.6em\slash$}}
\def \ttbar {\ensuremath{ \rm t \bar{t}             }}
\def\etmiss{\mbox{$\rm \Eslash_{T}\!$}}
\def \pt     {\ensuremath{ p_{\mathrm{T}}                       }}
\newcommand{\etmissscal}{\mbox{$\Eslash_T^{\rm{Scaled}}$ }}
\newcommand{\etmisssc}{\mbox{\ensuremath{\rm \etmiss^{\kern-0.6em\text{Scaled}}\!}       }}

\begin{document}

\title{Search for $H\rightarrow WW \rightarrow \ell \ell$ at  D\O} 

%

\author{Bj\"orn Penning}
\affiliation{University of Freiburg}

\begin{abstract}
A search for the Higgs boson in $H \rightarrow W W^* \rightarrow \ell \ell$ ($\ell = e, \mu$)  decays in $p \overline{p}$ collisions at a center-of-mass energy of $\sqrt{s}=1.96$ TeV is presented. The data have been collected by the Run II D\O\ detector between April 2002 and June 2008, corresponding to 3 fb$^{-1}$ of integrated luminosity. In order to maximize the sensitivity the multivariate technique of Artificial Neural Networks is used. No excess above the Standard Model background is observed and limits on the production cross section times branching ratio $\sigma \times BR(H \to WW^* \to \ell \ell)$ for $m_{\rm{H}}=115 - 200$ GeV using a Higgs mass grid of 5 GeV have been set. The upper limit as ratio of the SM prediction at 95\% C.L. for a Higgs boson mass of 165 GeV/$c^2$ is
\centering $\sigma (p\bar{p} \rightarrow) \times BR(H \rightarrow W^{+} W^{-})\, [{\rm (\sigma \times BR) / SM}] \le 1.9 ~\ (\le 2.0 ~\ {\rm expected}). $  
\end{abstract}

\maketitle

\thispagestyle{fancy}


\section{Introduction} 
A search for the Higgs boson decaying to the $WW^*$ final state is presented. The dileptonic decay modes  $H \rightarrow WW \rightarrow \ell \ell$  have been studied  using an integrated luminosity of about $\sim 3$ fb$^{-1}$ of RunII data recorded between 2002 and 2008. These decay modes provide the best sensitivity to a Standard Model (SM) Higgs boson search at the Tevatron at a mass of $m_{\rm H} \sim 160 ~GeV/c^2$. In order to maximize the signal to background separation  multivariate techniques are used. If combined with searches exploiting the $WH$ and $ZH$ associated production, these decay modes increase the sensitivity for the Higgs boson searches across a large mass range.
\section{Event Selection \label{sec:sel}}
\begin{figure}[htp]
  \begin{minipage}{0.32\textwidth}
    \centering 
    \includegraphics[width=1\textwidth]{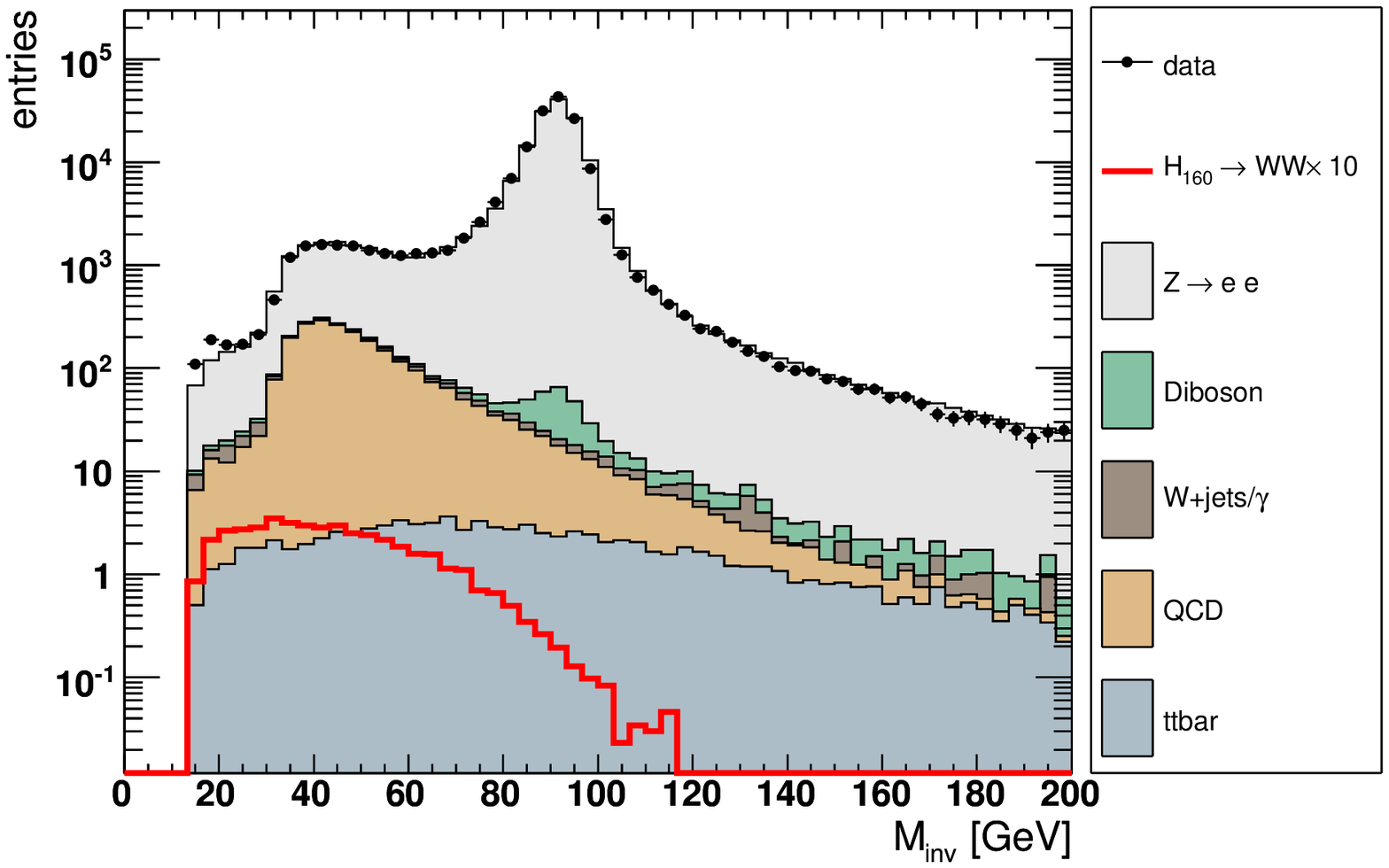}\\
    $ee$ channel
  \end{minipage}
  \begin{minipage}{0.32\textwidth}
      \centering 
    \includegraphics[width=1\textwidth]{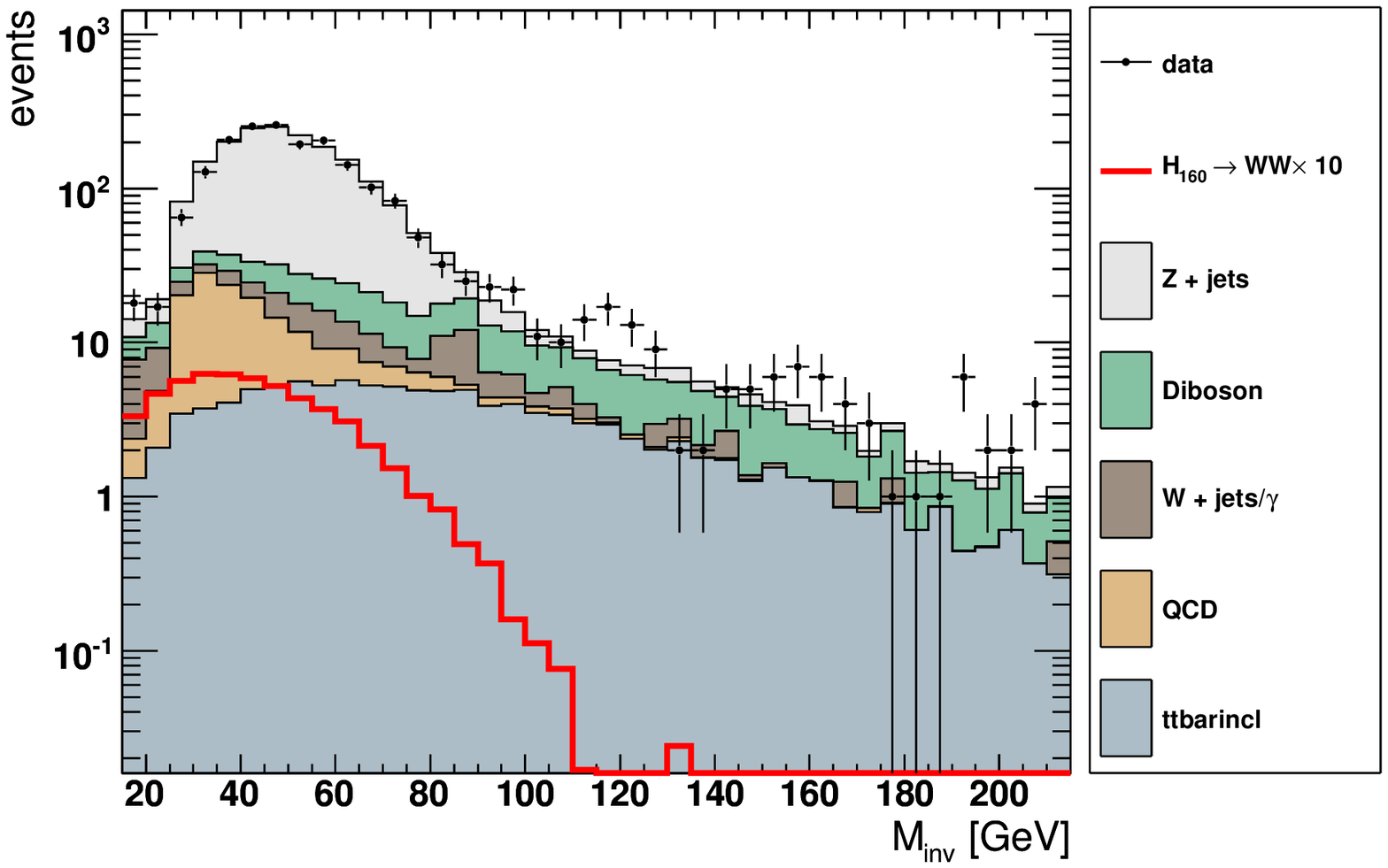}\\
    $e\mu$ channel
  \end{minipage} 
  \begin{minipage}{0.32\textwidth}
      \centering 
    \includegraphics[width=1\textwidth]{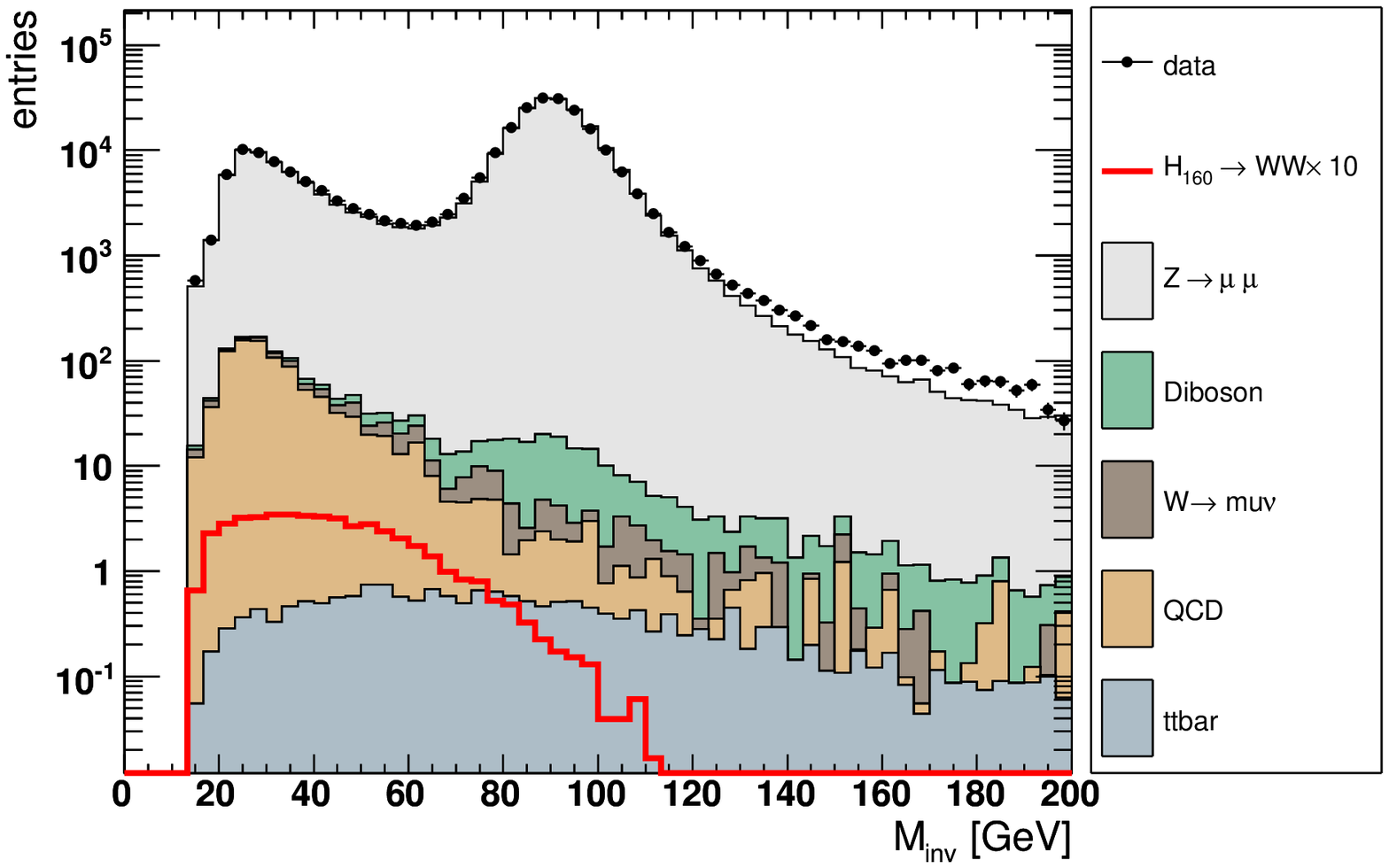}\\
    $\mu\mu$ channel
  \end{minipage} 
  \caption{ Invaraiant mass of the di-lepton system for the $ee$, $e\mu$ and $\muþmu$ channel respectively \label{fig_M} }
\end{figure}

In this channel we utilize that high mass Higgs boson decay dominantly via a pair of $W$ boson. Each $W$ boson is subsequently required to decay leptonically, producing a lepton and a neutrino escaping the detector. Since either electrons or muons are selected three different final states exist, $ee$, $e \mu$ and $\mu \mu$. Each channel features a distinctive signature characterized by two leptons, missing transverse energy ($\etmiss$) and little jet activity. The lepton candidates are selected by using a three level trigger system triggering on either single or di-lepton events. In the offline analysis electrons are identified using calorimeter and tracking information. Muons are reconstructed from hits in the wire chambers and scintillators in the muon system and must match a track in the central tracker. All leptons are required to be isolated. In all final states the leptons have to be oppositely charged. Muons are required to exceed a transverse momentum of at least 10 GeV  and the electrons of $15$ GeV. Additionally the invariant mass of the two leptons in each event is required to be greater than 15 GeV. This stage is referred to as the preselection stage. The invariant mass peak at this stage for each of the three channels is shown in Figure \ref{fig_M}.
Subsequently most of the multi-jet ("$QCD$'') background is removed by applying requirements on the missing transverse energy, $\etmiss$, and the scaled missing transverse energy ,\etmissscal, which is the $\etmiss$ divided by the $\etmiss$ resolution. This quantity is particularly sensitive to events where the missing energy could be a result of mismeasurements of jet energies in the transverse plane. A requirement on the minimal transverse mass, $M_T^{min}=\sqrt{2 \cdot \etmiss \cdot p_T^l \cdot (1-\cos (\Delta \phi))}$,  between one of the leptons and $\etmiss$ further reduces the various background processes. A large fraction of remaining back-to-back  $\rm Z/\gamma^{*} \rightarrow \ell \ell$ is reduced by rejecting events with a wide opening angle between the leptons.
\begin{table}[htb]
  \begin{center}
    \caption{\label{yields} Expected and observed number of events in each channel after preselection and final selections (the NN input stage). Statistical uncertainties in the expected yields are shown for all backgrounds whereas the systematic uncertainty is shown for the multi-jet background.}
    \begin{tabular}{|c|c|c|c|c|c|c|}
      \hline 
      &\textbf{$e\mu$ preselection} & \textbf{$e\mu$ final}& \textbf{$ee$ preselection} & \textbf{$ee$ final} &\textbf{$\mu\mu$ preselection} & \textbf{$\mu\mu$ final} \\
      \hline                                        
      $Z\to ee$       & $209.0\pm3.0$& $0.72\pm0.16$& $160463 \pm264$        & $73.6 \pm 5.1$      & $-$ & $-$   \\
      $Z\to \mu\mu$   & $151.1\pm0.6$& $2.14\pm0.06$& -                      & -                   & $256432\pm 230$ & $ 957\pm 14$ \\
      $Z\to \tau\tau$ & $2312\pm2$   & $2.45\pm0.05$& $835\pm8$              & $1.0 \pm0.3 $       & $1968  \pm 11$  & $ 5.5 \pm 0.5$ \\
      $\ttbar$        & $187.5\pm0.2$& $54.2\pm0.1$ & $96.9\pm0.2$           & $28.5\pm0.1$        & $19.4\pm0.1$    & $10.1 \pm 0.1$ \\
      $W+jets$        & $163.4\pm5.3$& $60.1\pm3.2$ & $174\pm7$              & $72.0\pm4.3$        & $149 \pm 3$     & $85.8 \pm 2.1$ \\
      $WW$            & $285.6\pm0.1$& $108.0\pm0.1$& $127.5\pm0.4$          & $45.7 \pm0.2 $      & $162.9 \pm 0.5$ & $91.3 \pm 0.3$ \\
      $WZ$            & $14.8\pm0.1$ & $4.9\pm0.1$  & $89.6\pm0.8 $          & $7.6 \pm0.2 $       & $51.6\pm0.5$    & $16.2\pm0.3$ \\
      $ZZ$            & $3.47\pm0.01$& $0.49\pm0.01$& $73.5 \pm0.3 $         & $5.4 \pm0.1 $       & $43.0  \pm 0.2$ & $13.5\pm0.1$ \\
      Multi-jet       & $190\pm168$  & $1\pm8$      & $2322\pm193$           & $4.3\pm8.3$         & $ 945\pm31$     & $63.6\pm8.0$ \\
      \hline Signal                                 
      ($m_H=160$ GeV) & $9.0 \pm0.1 $& $6.9 \pm0.1 $& $4.40\pm0.01$          & $3.49\pm0.01$       & $4.7 \pm0.1 $   & $4.09\pm0.06$ \\
      \hline                                        
      Total Background& $3516\pm168$ & $234\pm9$    & $164181\pm327$         & $238\pm11$         & $259770\pm232$   & $1242\pm16$ \\
      \hline                                        
      Data            & $3706$       & $234$        & $164290$               & $236$               & $263743$        & $1147$ \\
      \hline 
    \end{tabular}
  \end{center}
\end{table}

\section{Multivariate Discriminant \label{sec:multivar}}
\begin{figure}[htp]
  \begin{minipage}{0.32\textwidth}
    \centering 
    \includegraphics[width=1\textwidth]{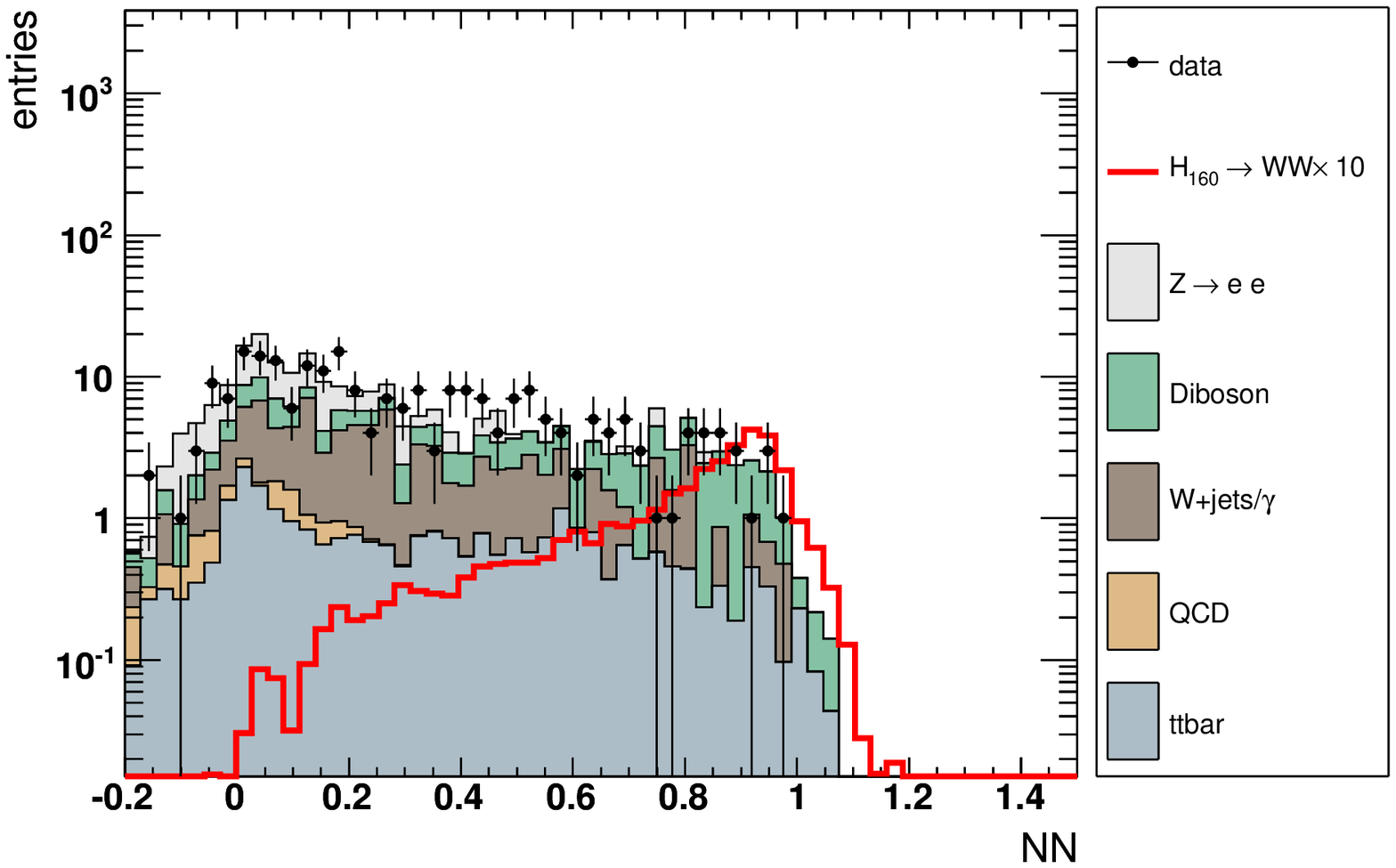}\\
    $ee$ channel
  \end{minipage}
  \begin{minipage}{0.32\textwidth}
      \centering 
    \includegraphics[width=1\textwidth]{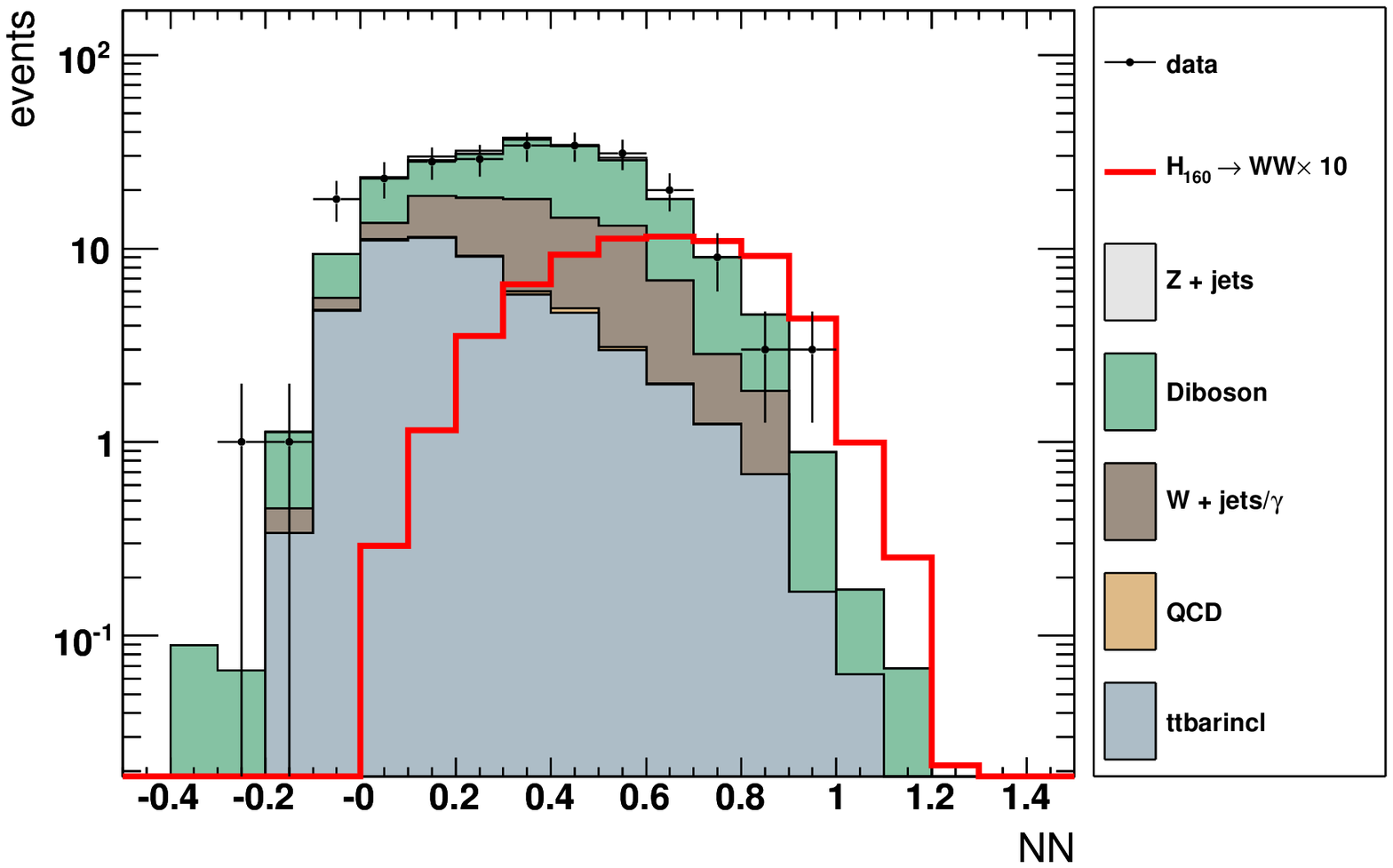}\\
    $e\mu$ channel
  \end{minipage} 
  \begin{minipage}{0.32\textwidth}
      \centering 
    \includegraphics[width=1\textwidth]{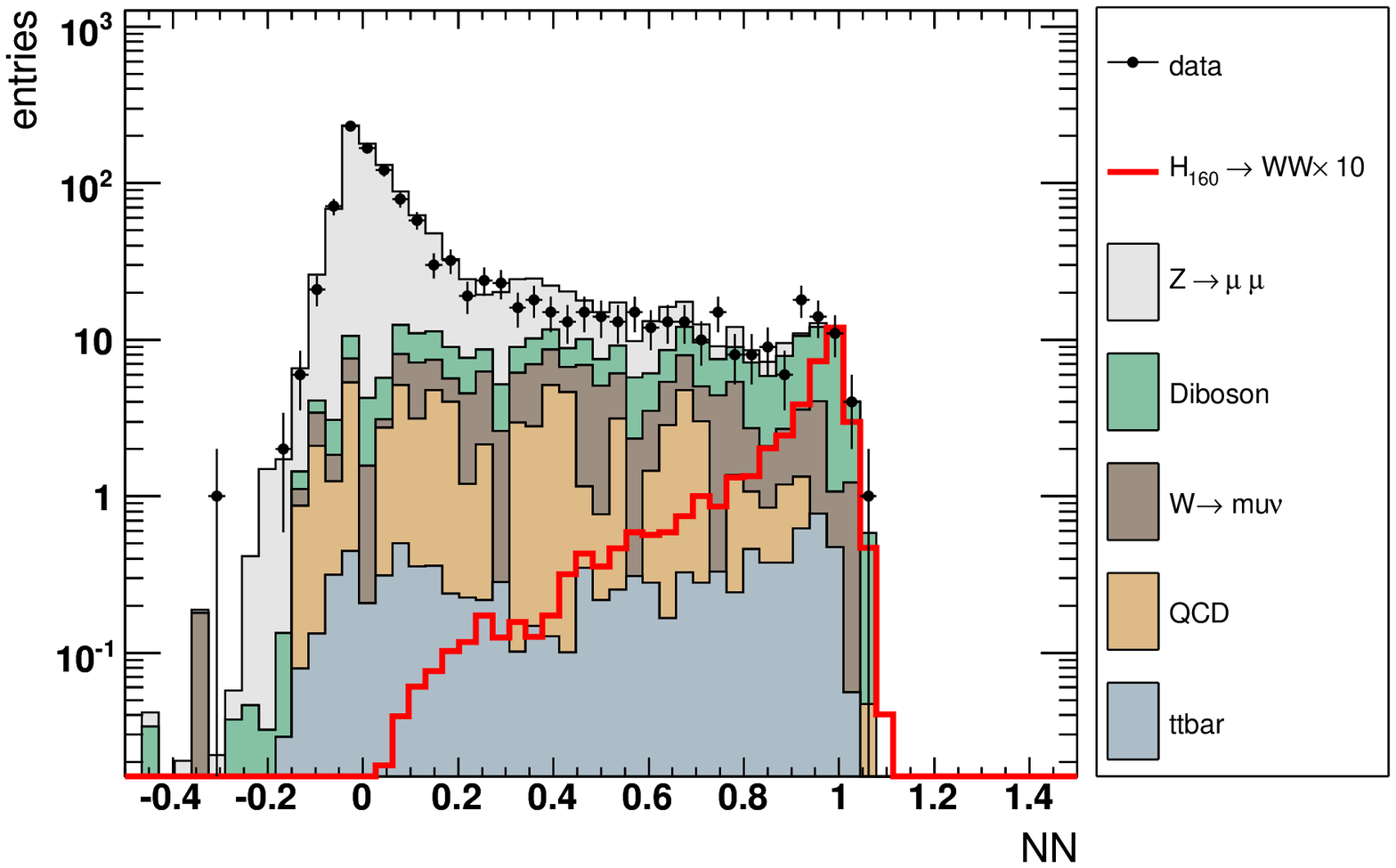}\\
    $\mu\mu$ channel
  \end{minipage} 
  \caption{ Neutral Network ouput distribution after final selection assuming a Higgs mass of $m_H=160$ GeV/c$^2$ for the $ee$, $e\mu$ and $\muþmu$ channel respectively \label{fig:NNoutput_ee} }
\end{figure}

To improve the separation of signal from background, artificial neural networks (NN) are used. Hereby signal like events get a high NN ouput value close to 1 assigned whereas background events peak around 0. The NNs were trained and tested using 50\% of the available background and signal samples, applying all MC/Data corrections like reconstruction and trigger efficiencies as in the analysis. The other half of the available samples has been used for the analysis, limit setting and Data/MC comparison. In order to exploit a potential signal as thoroughly as possible a separate NN has been trained for each Higgs mass and final state. The NN is based on a list of input variables which have been chosen based on their separation power between background and signal for the various distributions for each of the three channels. The NN is applied to all events passing the final selection requirements described in Section~\ref{sec:sel}. The NN output distribution for $m_{\rm H}=160$ GeV is displayed in Figure~\ref{fig:NNoutput_ee} for all three channels.
\section{Results and Conclusion}
\begin{figure}[htp]
  \begin{minipage}{0.49\textwidth}
    \centering 
    \includegraphics[width=0.65\textwidth]{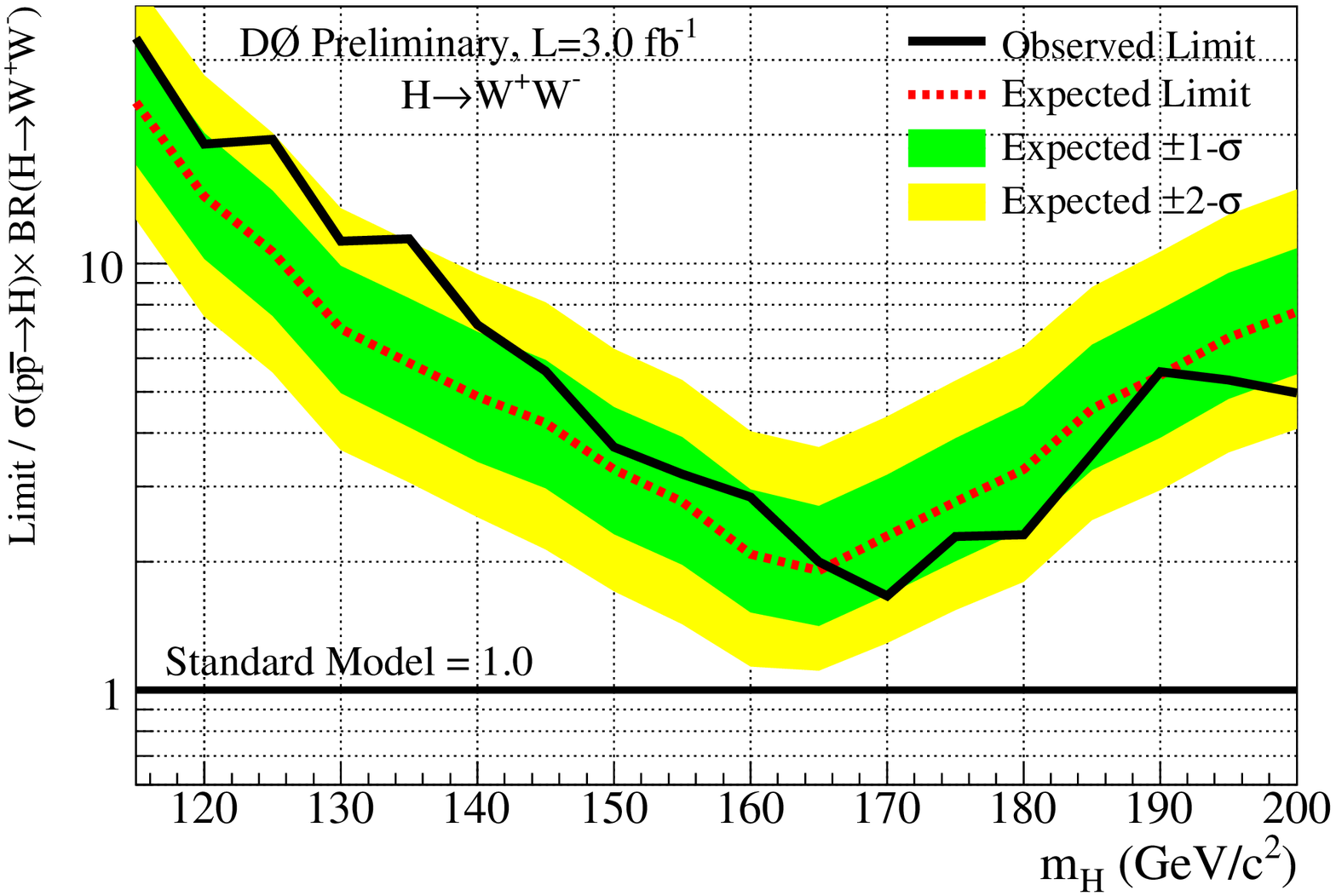}\\
    Expected and observed limit for $H \rightarrow WW \rightarrow \ell \ell$ 
  \end{minipage}
  \hfill
  \begin{minipage}{0.49\textwidth}
    \centering 
    \includegraphics[width=0.65\textwidth]{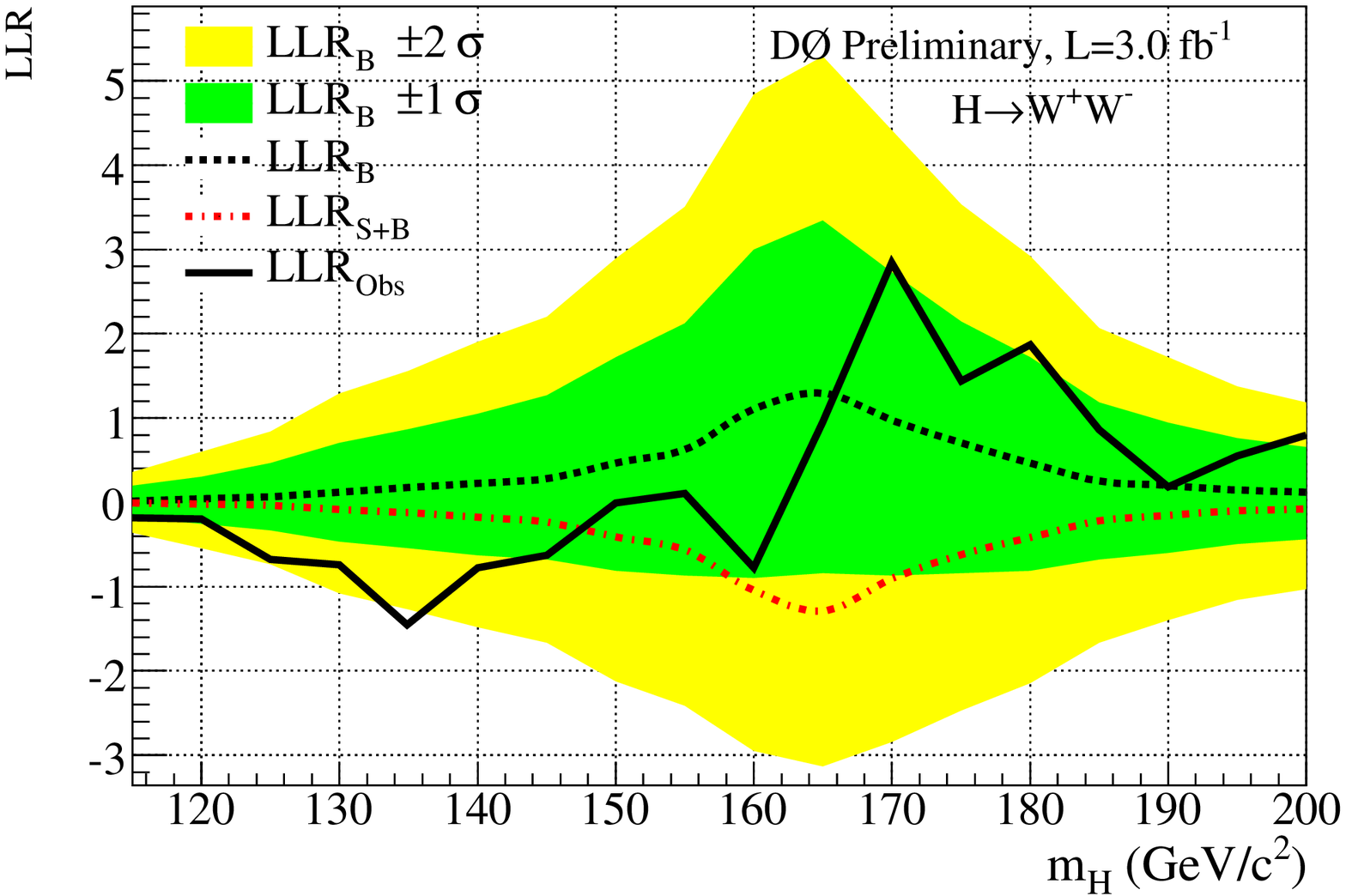}\\
    Expected and observed log likelihood for $H \rightarrow WW \rightarrow \ell \ell$ 
  \end{minipage} 
  \caption{ Excluded cross section times branching ratio $\sigma\times BR$($H \rightarrow W W^{(*)}$) at 95\% CL in units of the SM cross section (left) and LLR (right) for all three channels combined, using 3.0 fb$^{-1}$ of RunII data.  \label{fig_excl}}
\end{figure}

Limits on the cross section for Higgs boson production times the branching fraction into the $H \rightarrow W W^* \rightarrow \ell \ell$ final states are derived at the 95\% Confidence Level (CL). The estimates for the expected number of background and signal events depend on numerous factors, each of them introducing a systematic uncertainty. Two different kinds of systematics have been considered: flat systematics which show no dependency in terms of the NN value and shape systematics that modify the shape of the distribution of the NN variable used for limit setting. The following flat systematics have been studied: lepton reconstruction efficiencies (2.5-8\%), lepton momentum calibration (2\%), theoretical cross section (di-boson 7\%, \ttbar\ 10\%, $W$+jet 20\%), different modeling of multi-jet background.  The following systematics are implemented as shape dependent systematics: jet reconstruction efficiency(6\%), jet energy scale calibration (7\%), jet energy resolution (0.3\%), modeling of the instantaneous luminosity (0.3\%), modeling of the $z$-vertex distribution (1\%), modeling of $\pt(WW)$, $\pt(H)$, and $\pt(Z)$ (2-33\%). The systematic uncertainty on these \pt\ modeling has been determined by comparing the \pt\ distributions of \textsc{Pythia}, \textsc{Sherpa}, and \textsc{MC@NLO}. \textsc{Sherpa} and \textsc{MC@NLO} agree well with each other and generate harder \pt\ spectra than \textsc{Pythia}.  Since the luminosity is determined by normalizing to the $Z$-peak the systematic uncertainty is mainly a combination of the PDF uncertainty, uncertainty of the NNLO $Z$ cross section (4\%) and data/MC normalization factors (2\%). The total uncertainty on the background level is approximately 10\% and for the signal efficiency it is 9\%. 
\begin{table}[ht]
  \begin{center}
    \caption{\label{tab:alllimitbay} Expected and observed upper limits at 95\% CL for $\sigma \times BR (H \rightarrow W W^{(*)}) $ relative to the SM in RunII for different Higgs boson masses ($m_H$).}
    \begin{tabular}{|c|c|c|c|c|c|c|c|c|c|c|c|c|c|c|c|c|c|c|}
      \hline 
      $\mathbf{m_H}$\textbf{=} & \textbf{115} & \textbf{120} & \textbf{125} & \textbf{130} & \textbf{135} & \textbf{ 140 }& \textbf{145} & \textbf{150} & \textbf{155} & \textbf{160} & \textbf{165} & \textbf{170}& \textbf{175}& \textbf{180}& \textbf{185}& \textbf{190}& \textbf{195}& \textbf{200} \\
      \hline
      \textbf{Run II (exp.)} & 24    & 14    & 11    & 7.0   & 5.8   & 4.9   & 4.2   & 3.3   & 2.8   & 2.1   & 1.9   & 2.3   & 2.8   & 3.3   & 4.6   & 5.5   & 6.7   & 7.7   \\
      \textbf{Run II (obs.)} & 34    & 19    & 20    & 11    & 11    & 7.2  &  5.6  &  3.7  &  3.2  &  2.8  &  2.0  &  1.7  &  2.3  &  2.3  &  3.6  &  5.6  &  5.3  &  5.0  \\
      \hline
    \end{tabular}
  \end{center}
\end{table}
After all selection cuts the the expected background and data agrees well. Thus the NN output distributions are used to set limits on the production cross section times branching ratio $\sigma \times BR(H \rightarrow W W^{(*)})$. We calculate limits for each channel and all three channels combined, using a modified frequentist method, the CLs method, with a log-likelihood ratio (LLR) test statistic.  To minimize the degrading effects of systematics on the search sensitivity, the individual background contributions are fitted to the data observation by maximizing a profile likelihood function for each hypothesis. Table~\ref{tab:alllimitbay} and Figure \ref{fig_excl} present expected and observed upper limits at 95\% CL for $\sigma \times BR(H \rightarrow WW^{(*)})$ relative to that expected in the SM for each of the three final states and for their combination for each Higgs boson mass considered. 
In summary a search has been performed for the $H\rightarrow WW\rightarrow \ell \ell$ decay signature of the Standard  Model Higgs boson in
leptonic channels, using data corresponding to an integrated luminosity of $\approx$ 3 fb$^{-1}$.  No evidence for the Higgs particle is observed and these data have been combined with the CDF data. These combination lead to the first direct Standard Model Higgs excluison at Tevatron.


%

\end{document}